\documentclass[twocolumn,backref, breaklinks, colorlinks]{aastex63}
\hypersetup{linkcolor=blue,citecolor=blue,filecolor=cyan,urlcolor=magenta}

\usepackage{tabularx}
\usepackage{url}
\usepackage{graphicx}
\usepackage[T1]{fontenc}
\usepackage{ae,aecompl}
\usepackage{booktabs}
\usepackage{natbib}
\usepackage{longtable}

\usepackage{blindtext}
\usepackage{longtable}
\usepackage{tabu}
\usepackage{lineno}

\def\unit #1{\,{\rm #1}}

\newcommand\kev{\rm \,\unit{keV}}
\newcommand\mev{\rm \,\unit{MeV}}
\newcommand\gev{\rm \,\unit{GeV}}

\newcommand\ergs{\rm \,\unit{erg\,s^{-1}}}

\newcommand\funit{\rm \,erg\,cm^{-2}\,s^{-1}}
\newcommand\flu{\rm \,erg\,cm^{-2}}

\newcommand\lunit{\rm \,erg \,s^{-1}}

\newcommand\frb{\rm FRB~180916}

\newcommand\mpc{\unit{Mpc}}

\newcommand\chandra{{\it Chandra}}

\newcommand\swift{{\it Swift}}
\newcommand\ngswift{{\it The Neil Gehrels Swift observatory}}

\def\aap{A\&A} \def\apjl{ApJ}  
 \def\apjs{ApJS}

 \submitjournal{ApJ}

\shorttitle{FRB 180916}

\begin{document}


\title{ Limits on the hard X-ray emission from the periodic fast radio burst FRB 180916.J0158+65}


\author[0000-0003-2714-0487]{Sibasish Laha}

\affiliation{Center for Space Science and Technology, University of Maryland Baltimore County, 1000 Hilltop Circle, Baltimore, MD 21250, USA.}
\affiliation{Astrophysics Science Division, NASA Goddard Space Flight Center, Greenbelt, MD 20771, USA.}
\affiliation{Center for Research and Exploration in Space Science and Technology, NASA/GSFC, Greenbelt, Maryland 20771, USA}

\author[0000-0003-2714-0487]{Zorawar Wadiasingh}
\affiliation{Department of Astronomy, University of Maryland, College Park, Maryland 20742, USA}
\affiliation{Astrophysics Science Division, NASA Goddard Space Flight Center, Greenbelt, MD 20771, USA.}
\affiliation{Center for Research and Exploration in Space Science and Technology, NASA/GSFC, Greenbelt, Maryland 20771, USA}

\author[0000-0002-4299-2517]{Tyler Parsotan}
\affiliation{Center for Space Science and Technology, University of Maryland Baltimore County, 1000 Hilltop Circle, Baltimore, MD 21250, USA.}
\affiliation{Astrophysics Science Division, NASA Goddard Space Flight Center, Greenbelt, MD 20771, USA.}
\affiliation{Center for Research and Exploration in Space Science and Technology, NASA/GSFC, Greenbelt, Maryland 20771, USA}

\author[0000-0000-0000-0000]{Amy Lien}
\affiliation{University of Tampa, Department of Chemistry, Biochemistry, and Physics, 401 W. Kennedy Blvd, Tampa, FL 33606, USA}

\author[0000-0000-0000-0000]{George Younes} 
\affiliation{Astrophysics Science Division, NASA Goddard Space Flight Center, Greenbelt, MD 20771, USA.}

\author[0000-0000-0000-0000]{Bing Zhang}
\affiliation{Nevada Center for Astrophysics, University of Nevada, Las Vegas, NV 89154, USA}
\affiliation{Department of Physics and Astronomy, University of Nevada Las Vegas, Las Vegas, NV 89154 USA.}

\author[0000-0003-1673-970X]{S. Bradley Cenko}
\affiliation{Astrophysics Science Division, NASA Goddard Space Flight Center, Greenbelt, MD 20771, USA.}
\affiliation{Joint Space-Science Institute, University of Maryland, College Park, MD 20742, USA}

\author[0000-0002-1869-7817]{Eleonora Troja}
\affiliation{Astrophysics Science Division, NASA Goddard Space Flight Center, Greenbelt, MD 20771, USA.}
\affiliation{University of Rome Tor Vergata, Department of Physics, via della Ricerca Scientifica 1, 00100, Rome, IT}

\author[0000-0000-0000-0000]{Samantha Oates}
\affiliation{School of Physics and Astronomy, University of Birmingham, Birmingham B15 2TT, UK}
\affiliation{Institute for Gravitational Wave Astronomy, University of Birmingham, Birmingham B15 2TT, UK}

\author[0000-0000-0000-0000]{Matt Nicholl}
\affiliation{School of Physics and Astronomy, University of Birmingham, Birmingham B15 2TT, UK}
\affiliation{Institute for Gravitational Wave Astronomy, University of Birmingham, Birmingham B15 2TT, UK}

\author[0000-0000-0000-0000]{Eileen Meyer}
\affiliation{Department of Physics, University of Maryland, Baltimore County, 1000 Hilltop Circle, Baltimore, MD 21250, USA.}

\author[0000-0000-0000-0000]{Josefa Gonz\'alez ~Becerra}
\affiliation{Instituto de Astrof\'isica de Canarias (IAC), E-38200 La Laguna, Tenerife, Spain}
\affiliation{Universidad de La Laguna (ULL), Departamento de Astrof\'isica, E-38206 La Laguna, Tenerife, Spain}

\author[0000-0003-4790-2653]{Ritesh Ghosh}
\affiliation{Inter-University Centre for Astronomy and Astrophysics (IUCAA), Pune, 411007, India.}

\author[0000-0000-0000-0000]{Noel Klingler}
\affiliation{Center for Space Science and Technology, University of Maryland Baltimore County, 1000 Hilltop Circle, Baltimore, MD 21250, USA.}
\affiliation{Astrophysics Science Division, NASA Goddard Space Flight Center, Greenbelt, MD 20771, USA.}
\affiliation{Center for Research and Exploration in Space Science and Technology, NASA/GSFC, Greenbelt, Maryland 20771, USA}

\correspondingauthor{Sibasish Laha}
\email{sibasish.laha@nasa.gov,sib.laha@gmail.com}


\begin{abstract}
	
FRB 180916.J0158+65 is one of the nearest, periodically repeating, and actively bursting fast radio burst (FRB) which has been localized to the outskirts of a spiral galaxy. In this work we study the FRB with the hard X-ray $14-195\kev$ data from the Burst Alert Telescope (BAT) on board \ngswift{}. BAT uses coded mask technology giving a localization of $\lesssim 3$ arc-minute in the hard X-ray band, along with an accurate background estimation. BAT has been observing the source location in survey mode since February 2020. The survey mode observations involves background subtracted spectra, integrated over a time span ranging $300-2000$ seconds, at the source location (from Feb 2020-Jan 2022). We analyzed all the $\sim 230$ survey mode observations from BAT and checked for any signal in any of the observations. We did not detect any signal at $>5\sigma$ confidence level in any of the observations. We could estimate a $5\sigma$ upper limit on the $14-195\kev$ flux, which ranged between $4.5\times 10^{-10} - 7.6\times 10^{-9}\funit$. At the source distance this relates to a $5\sigma$ upper limit on luminosity of $5.08\times 10^{44}- 8.5\times 10^{45}\lunit$. With this estimate, we could rule out any persistent X-ray emission, at the source location for these snapshots of BAT observations. 

\end{abstract}



\section{Introduction}\label{sec:intro}

Fast radio bursts (FRBs) are bright, millisecond duration radio outbursts whose origin is still not clearly understood. To date, most FRBs possess dispersion measures (DM) larger than that of our own galaxy implying distances consistent with an extra-galactic origin. In a few cases the host galaxies of these FRBs have been localized. For e.g., FRB~121102 \citep{Tendulkar2017}, where the authors have found that it's located within a dwarf galaxy at a redshift of $z\sim 0.192$ and there is a faint persistent radio source of unknown origin at the FRB location \citep{Chatterjee2017}. Most FRBs have not been found to repeat. But  deeper sky surveys such as CHIME/FRB are revealing an increasing number of repeater FRBs \citep{fonseca2020,chime2019,chime_catalog_2021}. It has been proposed that most FRBs that we observe are actually repeaters because their all sky rates largely exceed the all sky rates of known cataclysmic events \citep[see e.g.,][and references therein]{Nicholl+17,Ravi2019,Zhang2020}{}.


\frb{} is one of the most well known repeaters and also one of the most nearby FRBs detected to date. In December 2019, the CHIME/FRB collaboration reported the discovery of eight new repeating FRBs including \frb{}, which was localized to a star forming region of a nearby massive spiral galaxy at a redshift of $z=0.0337\pm 0.0002$ \citep{Marcote2020}. The \frb{} exhibited $\sim 38$ radio bursts during September 2019- October 2020, showing a clear periodicity of $16.35\pm 0.15$ days, the first for any FRB. All the bursts arrived in a 5 day phase window and almost $50\%$ of the bursts arrived in the 0.6 day phase window, and it's the most active FRB in the published CHIME/FRB sample \citep{CHIME+19REPEATERS}. The periodicity suggests a mechanism which periodically modulates the observed emission itself, or a periodic burst, or something else. Hence, this source is the focus of intense multiwavelength campaigns.


High energy emission from magnetars has long been thought to be one of the important mechanisms by which the FRBs may also be generated \citep[see][and references therein]{Margalit2020}. The recent discovery of the connection between FRB200428 and SGR~1935 \citep{Mereghetti2020,Li21,Younes2020,Younes2021,Margalit2020,CHIME2020_SGR1935,Stare2_2021_SGR1935} confirmed this fact, that at least some of the FRBs are produced by magnetar short bursts. The typical radio to X-ray fluence measured was $\eta \equiv E_{\rm R}/E_{\rm X}\sim 10^{-5}$, implying that the radio emission is weak and contributes little to the overall burst energy budget. Moreover the radio fluence of FRB200428 was somewhat lower compared to its extragalactic counterparts (the STARE2 burst energetics however are strictly a lower limit since some of the burst appears to be out-of-band). Variable high energy emission at different timescales (ms to years) is one of the key features of magnetars, and may have a relation to the FRB emission \citep{coti2018,Younes2020,Younes2015,Palmer2005}.

Here we discuss three most common types of emission phenomenology from magnetars. Firstly, the flares associated with the short bursts which can be bright ($L_{\rm X}\le 10^{42}\lunit{}$) and lasts a few 100s of ms. These short bursts may happen in isolation, or during a burst storm when hundreds of them are detected continuing for days. Secondly, magnetars can exhibit giant flares, which consist of a very bright pulse ($L_{\rm X}\le 10^{47}\lunit{}$) at sub-ms level, followed by a softer pulsating tail lasting for several minutes. These events have been detected in three Galactic magnetars \citep[e.g., SGR1806 ][]{Mazets1979,Hurley1999,Palmer2005}{}. Thirdly, on longer timescales, magnetars recurrently enter outburst phases, when their persistent flux level increases by orders of magnitude, accompanied by random spectral and temporal variability. These phases often persist for a few months to years, after which they recover to their pre-outburst levels. In this work we focus on detecting the second and third kinds of emission from a putative extragalactic magnetar powering \frb{}. \swift{} Burst Alert Telescope (BAT) survey observations, which cover the energy range $14-195\kev$ and having a time-integration of a few 100s of seconds, are constraining to models of young magnetars undergoing giant flares.


Several works studying the X-rays to $\gamma$-rays have put upper limits on the possible non-detection from the sky location of the \frb{}. \cite{Verrecchia2021} using AGILE in the energy range $0.4\mev{}-30\gev{}$ could put an upper limit on the total hard X-ray energy of $\sim 10^{46}\ergs{}$ comparable to that of SGR~1806-20. Although the extragalactic FRBs emit radio pulses of energies which are significantly larger than that detected from SGR~1935, there are no simultaneous or contemporaneous detections of $\kev,\, \mev,\, \gev$ photons to-date. Contemporaneous X-ray and radio observations of the \frb{} were carried out by \citet{Scholz2020}, with CHIME and \chandra{} X-ray observatory. They detected no X-ray events in excess of the background, and the non-detections imply a $5\sigma$ fluence limit of $<5\times10^{-10}\flu{}$ in the $0.5-10\kev$ energy range during the prompt emission and $<1.3\times10^{-9}\flu{}$ any time during the observations. Given the cosmological distance of this FRB, these relate to the total isotropic energies $<1.6\times 10^{45}$ erg and $<4\times10^{45}$ erg respectively. In the $10-100\kev$ Fermi GBM search for prompt emission \citep{Scholz2020} led to a fluence upper limit of $9\times 10^{-9}\flu$. \cite{Tavani2020} searched for contemporaneous $\gamma$-ray emission from \frb{} using {\it AGILE} and X-ray emission using XRT telescope onboard \ngswift{} (\swift{} from now on), during the bursting phases of the source in February and March 2020. They did not detect any hard X-ray or $\gamma$-ray photons, and could provide a persistent flux upper limit of $5\times 10^{-14}\funit$ which translates to an isotropic luminosity of $L_{\rm X}\sim 1.5\times 10^{41}\lunit$. \cite{Guidorzi2020} using the Insight-Hard X-ray Modulation Telescope (Insight-HXMT) could provide a prompt emission upper limit in hard X-rays ($1-100\kev$) of $<10^{46}$ erg over a timescale of $\Delta t<0.1$s. They could rule out giant flares similar to the ones that were observed in Galactic magnetars.

FRB sources like magnetars may emit the majority of their flux at photon energies $>10\kev$ \citep{Margalit2020,Scholz2017,turolla2015,woods2006}. To study the possible hard X-ray emission in the $14-195\kev$ energy band, we analyze all the available Survey mode observations from BAT onboard \swift{}. BAT's well-suited energy range ($14-195 \kev$) coupled with hundreds of pointing on the source over a period extending Feb 2020- Jan 2022, serves as a unique opportunity for us to capture any hard X-ray emission from the \frb{} and/or put upper limits on the short term ($100$s of sec) or long term ($\sim$year) persistent X-ray emission from the source.

The paper is arranged as follows: Section \ref{Sec:obs} discusses the observations and data analysis. Section \ref{Sec:discussion} discusses the results, followed by conclusions in Section \ref{Sec:conclusions}.

\section{Observation and data analysis}\label{Sec:obs}

\frb{} has been observed by {\it Swift}  BAT \citep{gehrels2004,barthelmy2005} $\sim 220$ times from Feb 2020 - Jan 2022 in the survey mode. We checked in the CHIME FRB catalog \citep{chime_catalog_2021} for the time stamps of the $\sim 90$ radio bursts detected to date from FRB~180916. Unfortunately, none of the CHIME reported burst times were coincident with our BAT survey observations. We also note that there are no time tagged event (TTE) data avaiable in BAT archive where the FRB~180916 is in the field-of-view of BAT. For out of the field-of-view sources, the response matrix of BAT is highly uncertain and hence any measurement of flux is associated with large errors, hence we did not use them in this work. Below we describe the methods used to obtain the survey datasets, reprocess and analyze them. We have extensively used {\it bat-tools} from HEASOFT.

The BAT ``survey" mode data are also known as detector plane histograms or DPHs. As opposed to event data, BAT survey data is accumulated in histograms on-board the spacecraft, with typical integration times of 300 seconds or more. An 80-channel binned spectrum is recorded for each of the active detectors and saved in the DPH files. In this work we use all the available BAT datasets and we explain below the steps we have taken to reprocess and analyze them (See Table \ref{Table:obs} for a full list of observations). 

The BAT survey datasets were downloaded from HEASARC, and every survey observation has one or more pointings and hence different integration time. The task {\it batsurvey}  performs basic analysis of BAT survey data and reduces a set of ``raw'' observed DPHs. Most importantly, it performs data screening that the BAT team has found vital for obtaining good quality results. It produces sky images and source fluxes for each independent ``snapshot," corresponding to a single pointed visit by BAT. We chose a set of 8 independent energy bins, and {\it batsurvey} recorded the images and fluxes in each of those bands separately. {\it batsurvey} operates on a single BAT observation. For multiple observations, we used {\it batsurvey} once for each observation.  We provided the source RA and Dec in the catalog which was used as an input to {\it batsurvey}. 

The output from batsurvey gives us exposure-specific results. For example, one of the important files produced for each pointing is the flux catalog. This catalog contains sources listed in the input catalog (which in this case includes \frb{}) as well as sources detected by blind search. From this file, we create the lightcurve for the source \frb{} using the command ``batsurvey-catmux". We then extract the Tstart, and $\Delta T$ from the lightcurve and obtain the time integrated spectrum with eight energy channels. 

In the first round, to generate the source spectrum we choose ``CENT-RATE" as the column for rate array, and use the background variance (BKG-VAR) for the error on the rate. We note that the source spectrum is already background subtracted due to coded mask technique, hence sometimes when the background variance is larger than the ``net source" counts, then the count rate may become negative. Once the spectrum is obtained we obtain the response matrix using the task ``batdrmgen". We used the model {\tt cflux*powerlaw} in XSPEC notation to estimate the flux in the $14-195\kev$ energy band. We froze the powerlaw normalization to a value of $1e-03$ and kept the powerlaw slope ($\Gamma$) free, in order to calculate the flux using the model above. We did not detect any excess emission above the background (at a $>5\sigma$ confidence) in any of the pointings of this source.

In the second round, we re-extracted the spectrum for every pointing, now using the $5\times$ BKG-VAR as the rate array. Here we use a simple {\tt powerlaw} model with $\Gamma=1$ fixed, to estimate the flux in the $14-195\kev$ energy range. Note that this is the $5\sigma$ upper limit on the background variance and hence the $5\sigma$ upper limit on the flux. In Table 1 we quote the $5\sigma$ upper limit on the $14-195\kev$ flux obtained for all the available BAT pointings. To carry out all the above steps in a coherent way, we developed a user friendly pipeline in python, which will be published in the near future.


\section{Results and Discussion}\label{Sec:discussion}


We have carried out a hard X-ray counterpart search of the \frb{} using ``survey" mode observations of BAT onboard \swift{}. We have analyzed all available observations as of 15th January 2022, using standard analysis methods. We do not detect any source signal in excess of background with $>5\sigma$ significance. The $5\sigma$ upper limit on the background flux range from $4.5\times 10^{-10}- 7.6\times 10^{-9}\funit$ for different pointings of BAT survey observations. At the source distance this relates to a $5\sigma$ upper limit on luminosity of $5.08\times 10^{44}- 8.5\times 10^{45}\lunit$. Below we discuss our results in context to previous estimates of X-ray limits on this source and also discuss the possible progenitor scenarios. We note that the CHIME measured fluences for this FRB vary considerably, and one representative value is $\sim 27$ Jy-ms (for the burst ID:181222, in the CHIME catalog). Unfortunately, we did not have any simultaneous CHIME and BAT (in-field-of-view) observation.


 Observations with {\it AGILE} telescope \citep{tavani2021}, working in the energy range $400\kev-100\mev$, could constrain the bursting X-ray energy of \frb{} to $\sim 10^{46}$ erg, which is smaller than that observed from giant flares from Galactic magnetars. Similar upper limits of $\sim 10^{46}$ erg on the bursting energy of the \frb{} were obtained using {\it Insight-HMXT} in the energy range $1-100\kev$ \citep{Guidorzi2020}. On the soft X-ray range $0.3-10\kev$ strong limits on persistent luminosity ($<2\times 10^{40}\lunit$) and prompt emission fluence ($<5\times 10^{-10}\flu$) could be obtained during the bursting phases of the \frb{} using observations from \chandra{} \citep{Scholz2020}. Similarly, \cite{Tavani2020} could put a cumulative upper limit on the X-ray flux using \swift{} XRT observations $F<5\times 10^{-14}\funit$, obtained during the five day of active period of the source. These clearly rule out magnetar giant flares and/or SGR~1935 type events.  
 
  The \swift{}-BAT flux upper limits in $14-195\kev$, obtained in this work, of a few $\times 10^{-9}\funit$ correspond to luminosity limit of about $\sim 10^{45}\lunit$ at the source location (that is at $149\mpc$). We discuss the following three progenitor scenarios:
  
  \begin{itemize}
 \item Magnetar Giant Flares (MGF): Typical magnetar giant flares with luminosity $\sim 10^{47}\lunit$, lasting for $\sim 0.1$sec\citep{burns2021}, would lead to an average luminosity of $\sim 10^{43}\lunit$, when averaged over 1000 seconds of a typical BAT survey observation. This would correponds to a flux of $\sim 4\times 10^{-12}\funit$. This flux level is well below the flux level that BAT is sensitive to, and hence we cannot rule out an MGF event.

 \item  Pulsating tails of MGF: The pulsating tails have a typical luminosity of $10^{42}\lunit$, and their peak energy lies in the BAT energy range, and they extend for several seconds \citep{burns2021,Younes2020}. The corresponding flux emitted by such pulsating tails would be $\sim 10^{-13}\funit$ which is well below the flux level that BAT is sensitive to.

  \item For an SGR~1935 type event:  For such an event happening at the location of \frb{} the expected $20-200\kev$ flux is about a few$ \times 10^{-14}\funit$ \citep{Mereghetti2020,Li21}. This flux level is also well below the flux level that BAT is sensitive to, and hence we cannot rule out an event like SGR~1935.

  Our limits exclude any persistent emission or transient activity with the source position down to $\sim 10^{45}$ erg s$^{-1}$. The possible progenitor of \frb{} might be older than canonical magnetars, based on the offset reported in \cite{2021ApJ...908L..12T}. The authors suggest that the spatial offset and hence the timescale (of travel from its presumed birth site) points more towards high-mass X-ray binaries and gamma-ray binaries, rather than magnetars. Given the unpredictable nature of FRBs and magnetars, such multi-band long term monitoring snapshots are extremely useful. Future opportunities with prompt downlink and analysis of \swift{}-BAT time-tagged-event (TTE) data of FRBs \citep[GUANO,][]{aaron_guano_2020,aaron2021}, contemporaneous with radio bursts will open up new avenues of hard X-ray follow up of these hitherto unknown phenomenon.

\end{itemize}

\section{Conclusions}\label{Sec:conclusions}

We searched for high-energy transients in Swift BAT survey mode data associated with \frb{} from February 2020 to January 2022. We did not detect any significant emission in the $14-195\kev$ energy band in any of our observational snapshots when the FRB~180916 was in the field of view of BAT. The upper limits on the flux estimated in these observations exclude any persistent emission or transient activity with the source position down to $\sim 10^{45}$ erg s$^{-1}$, for all the snapshots. Our results confirm and corroborate previous limits of high energy transients by {\it Fermi-GBM}, {\it AGILE}, {\it HXMT-Insight} and {\it XMM-Newton}.

\clearpage

\begin{table}

{\footnotesize
	\caption{List of Swift-BAT survey-mode observations of FRB~180916, and the $5\sigma$ upper limit on the fluxes in the energy band $14-195\kev$. }\label{Table:obs}
	 
\begin{tabular}{lllllllllllllll} \hline\hline

Obs ID  &	 Pointing ID	&	Start time (MET) &	Start time (UTC) &	exposure &	flux upper-lim\\
		&number		&		&year:day:hr:min:sec &(sec)		&$\funit{}$	 \\ \hline \\

00013201002 &	20200341843 &	602448230.0 &	2020Feb03 at 18:43:25 &	240.0 	 &	3.35e-09 \\
           
00013201005 &	20200371529 &	602695795.0 &	2020Feb06 at 15:29:30 &	1088.0 	 &	2.32e-09 \\
           
00013201006 &	20200381645 &	602786919.0 &	2020Feb07 at 16:48:14 &	684.0 	 &	2.81e-09 \\
            &	20200381523 &	602781819.0 &	2020Feb07 at 15:23:14 &	1044.0 	 &	2.35e-09 \\
           
00013201007 &	20200391516 &	602867799.0 &	2020Feb08 at 15:16:14 &	1044.0 	 &	2.39e-09 \\
            &	20200391702 &	602874130.0 &	2020Feb08 at 17:01:45 &	233.0 	 &	4.99e-09 \\
           
00013201008 &	20200561709 &	604343363.0 &	2020Feb25 at 17:08:58 &	300.0 	 &     2.51e-09 \\
            &	20200561535 &	604337736.0 &	2020Feb25 at 15:35:11 &	179.0 	 &	1.62e-09 \\
            &	20200561400 &	604332006.0 &	2020Feb25 at 13:59:41 &	177.0 	 &	5.90e-09 \\
           
00013201009 &	20200651430 &	605111453.0 &	2020Mar05 at 14:30:28 &	600.0 	 &	4.00e-09 \\
           
00013201011 &	20200671419 &	605283823.0 &	2020Mar07 at 14:23:18 &	300.0 	 &	5.38e-09 \\
           
00013201014 &	20200701357 &	605541476.0 &	2020Mar10 at 13:57:31 &	1567.0 	 &	2.39e-09 \\
           
00013201019 &	20200871153 &	607002839.0 &	2020Mar27 at 11:53:34 &	1744.0 	 &	2.31e-09 \\
           
00013201021 &	20201141454 &	609346533.0 &	2020Apr23 at 14:55:08 &	150.0 	 &	8.09e-09 \\
            &	20201141435 &	609345333.0 &	2020Apr23 at 14:35:08 &	900.0 	 &	3.60e-09 \\
           
00013201023 &	20201161102 &	609505344.0 &	2020Apr25 at 11:01:59 &	1659.0 	 &	2.95e-09 \\
           
00013201024 &	20201171055 &	609591323.0 &	2020Apr26 at 10:54:58 &	1660.0 	 &	2.96e-09 \\
           
00013201025 &	20201181536 &	609694745.0 &	2020Apr27 at 15:38:40 &	1378.0 	 &	2.73e-09 \\
           
00013201026 &	20201191526 &	609780416.0 &	2020Apr28 at 15:26:31 &	1687.0 	 &	2.50e-09 \\
           
00013201027 &	20201201522 &	609866526.0 &	2020Apr29 at 15:21:41 &	1677.0 	 &	2.36e-09 \\
           
00013201028 &	20201470743 &	612171836.0 &	2020May26 at 07:43:30 &	727.0 	 &	2.92e-09 \\
            &	20201470915 &	612177309.0 &	2020May26 at 09:14:43 &	654.0 	 &	3.13e-09 \\
           
00013201029 &	20201480730 &	612257414.0 &	2020May27 at 07:29:48 &	1069.0 	 &	2.44e-09 \\
            &	20201480904 &	612263088.0 &	2020May27 at 09:04:22 &	855.0 	 &	2.91e-09 \\
           
00013201030 &	20201490859 &	612349197.0 &	2020May28 at 08:59:31 &	666.0 	 &	3.11e-09 \\
           
00013201031 &	20201500856 &	612435365.0 &	2020May29 at 08:55:39 &	478.0 	 &	3.61e-09 \\
           
00013201032 &	20201510852 &	612521583.0 &	2020May30 at 08:52:37 &	240.0 	 &	5.105e-09 \\
           
00013201034 &	20201630148 &	613532923.0 &	2020Jun11 at 01:48:17 &	560.0 	 &	3.52e-09 \\
            &	20201630943 &	613561411.0 &	2020Jun11 at 09:43:05 &	1112.0 	 &	3.087e-09 \\
           
00013201035 &	20201641554 &	613670061.0 &	2020Jun12 at 15:53:55 &	1302.0 	 &	2.42e-09 \\
           
00013201036 &	20201652211 &	613779079.0 &	2020Jun13 at 22:10:53 &	1484.0 	 &	2.237e-09 \\
           
00013201037 &	20201661402 &	613836136.0 &	2020Jun14 at 14:01:50 &	1727.0 	 &	2.16e-09 \\
           
00013201038 &	20201671357 &	613922236.0 &	2020Jun15 at 13:56:50 &	1607.0 	 &	2.183e-09 \\
           
00013201039 &	20201681350 &	614008228.0 &	2020Jun16 at 13:50:02 &	1595.0 	 &	2.11e-09 \\
            &	20201682335 &	614043344.0 &	2020Jun16 at 23:35:18 &	439.0 	 &	3.88e-09 \\
           
00013201040 &	20201692005 &	614117139.0 &	2020Jun17 at 20:05:13 &	1584.0 	 &	2.23e-09 \\
           
00013201041 &	20201700231 &	614140312.0 &	2020Jun18 at 02:31:26 &	1691.0 	 &	2.25e-09 \\
           
00013201043 &	20201800805 &	615024353.0 &	2020Jun28 at 08:05:27 &	600.0	 &	3.32e-09 \\
            &	20201800604 &	615017085.0 &	2020Jun28 at 06:04:19 &	1098.0 	 &	2.77e-09 \\
           
00013201044 &	20201810613 &	615103998.0 &	2020Jun29 at 06:12:52 &	1005.0 	 &	2.85e-09 \\
            &	20201810754 &	615110061.0 &	2020Jun29 at 07:53:55 &	762.0 	 &	3.03e-09 \\
           
00013201045 &	20201820727 &	615194850.0 &	2020Jun30 at 07:27:04 &	333.0 	 &	4.60e-09 \\
            &	20201820606 &	615189999.0 &	2020Jun30 at 06:06:13 &	1200.0	 &	2.49e-09 \\
           
00013201046 &	20201830410 &	615269416.0 &	2020Jul01 at 04:09:50 &	467.0 	 &	4.18e-09 \\
            &	20201830603 &	615276238.0 &	2020Jul01 at 06:03:32 &	965.0 	 &	2.75e-09 \\
           
00013201047 &	20201840559 &	615362370.0 &	2020Jul02 at 05:59:04 &	813.0 	 &	3.10-09 \\
           
00013201048 &	20201850553 &	615448424.0 &	2020Jul03 at 05:53:18 &	799.0 	 &	2.98e-09 \\
           
00013201049 &	20201860550 &	615534625.0 &	2020Jul04 at 05:49:59 &	578.0 	 &	8.3e-10 \\

\\ \hline \\
\end{tabular}

These datasets are all publicly available in the HEASARC archive as on 15th January 2022. \\
The start times of the observations are given in two units. MET: Mission elapsed time, and UTC: Universal Coordinated time.\\

}
\end{table}

\clearpage

\begin{table}
{\footnotesize
	\caption{List of Swift-BAT survey-mode observations of FRB~180916. (Continued) }\label{Table:obs}
	  \begin{tabular}{lllllllllllllll} \hline\hline

Obs ID  &	 Pointing ID	&	Start time (MET) &	Start time (UTC) &	exposure &	flux upper-lim\\
		&number		&		&year:day:hr:min:sec &(Sec)		&$\funit{}$	 \\ \hline \\

00013201051 &	20201960600 &	616399216.0 &	2020Jul14 at 05:59:50 &	1007.0 	 &	3.00e-09 \\
            &	20201960427 &	616393656.0 &	2020Jul14 at 04:27:10 &	687.0 	 &	3.42e-09 \\
           
00013201052 &	20201970549 &	616484971.0 &	2020Jul15 at 05:49:05 &	1652.0 	 &	2.42e-09 \\
           
00013201053 &	20201980551 &	616571487.0 &	2020Jul16 at 05:51:01 &	576.0 	 &	3.81e-09 \\
            &	20201980419 &	616565963.0 &	2020Jul16 at 04:18:57 &	460.0 	 &	4.17e-09 \\
           
00013201054 &	20201990539 &	616657157.0 &	2020Jul17 at 05:38:51 &	826.0 	 &	3.196e-09 \\
           
00013201055 &	20202000531 &	616743121.0 &	2020Jul18 at 05:31:35 &	842.0 	 &	3.33e-09 \\
            &	20202000356 &	616737379.0 &	2020Jul18 at 03:55:53 &	764.0 	 &	3.298e-09 \\
           
00013201057 &	20202020347 &	616909656.0 &	2020Jul20 at 03:47:10 &	387.0 	 &	4.45e-09 \\
            &	20202020523 &	616915431.0 &	2020Jul20 at 05:23:25 &	432.0 	 &	4.26e-09 \\
           
00013201058 &	20202130447 &	617863661.0 &	2020Jul31 at 04:47:15 &	802.0 	 &	3.74e-09 \\
            &	20202130311 &	617857887.0 &	2020Jul31 at 03:11:01 &	516.0 	 &	4.06e-09 \\
           
00013201059 &	20202140437 &	617949461.0 &	2020Aug01 at 04:37:15 &	1282.0 	 &	3.12e-09 \\
            &	20202140303 &	617943818.0 &	2020Aug01 at 03:03:12 &	505.0 	 &	4.40e-09 \\
           
00013201060 &	20202150430 &	618035428.0 &	2020Aug02 at 04:30:02 &	1200.0 	 &	3.23e-09 \\
            &	20202150255 &	618029745.0 &	2020Aug02 at 02:55:19 &	258.0 	 &	6.34e-09 \\
           
00013201061 &	20202160425 &	618121550.0 &	2020Aug03 at 04:25:24 &	900.0 	 &	3.718e-09 \\
           
00013201062 &	20202170416 &	618207380.0 &	2020Aug04 at 04:15:54 &	1200.0 	 &	3.0127e-09 \\
           
00013201063 &	20202180409 &	618293397.0 &	2020Aug05 at 04:09:31 &	1200.0 	 &	2.9025e-09 \\
           
00013201064 &	20202190402 &	618379338.0 &	2020Aug06 at 04:01:52 &	1200.0 	 &	2.627e-09 \\
           
00013201065 &	20202290301 &	619239722.0 &	2020Aug16 at 03:01:36 &	1500.0 	 &	2.2564e-09 \\
           
00013201066 &	20202300259 &	619325953.0 &	2020Aug17 at 02:58:47 &	300.0 	 &	1.94e-09 \\
            &	20202300313 &	619326853.0 &	2020Aug17 at 03:13:47 &	351.0 	 &	4.289e-09 \\
           
00013201067 &	20202310253 &	619412011.0 &	2020Aug18 at 02:53:05 &	1173.0 	 &	2.4877e-09 \\
           
00013201075 &	20202450243 &	620621040.0 &	2020Sep01 at 02:43:34 &	1023.0 	 &	2.7753e-09 \\
            &	20202450127 &	620616464.0 &	2020Sep01 at 01:27:18 &	799.0 	 &	3.1037e-09 \\
           
00013201076 &	20202460233 &	620706802.0 &	2020Sep02 at 02:32:56 &	1481.0 	 &	2.439e-09 \\
           
00013201077 &	20202470252 &	620794340.0 &	2020Sep03 at 02:51:54 &	223.0 	 &	5.8695e-09 \\
            &	20202470227 &	620792840.0 &	2020Sep03 at 02:26:54 &	600.0 	 &	3.7665e-09 \\
           
00013201079 &	20202490225 &	620965519.0 &	2020Sep05 at 02:24:53 &	524.0 	 &	3.803e-09 \\
            &	20202490215 &	620964919.0 &	2020Sep05 at 02:14:53 &	600.0 	 &	3.612e-09 \\
           
00013201080 &	20202500050 &	621046242.0 &	2020Sep06 at 00:50:16 &	862.0 	 &	1.07e-09 \\
            &	20202500210 &	621051011.0 &	2020Sep06 at 02:09:45 &	300.0 	 &	1.14e-09 \\
           
00013201081 &	20202510044 &	621132260.0 &	2020Sep07 at 00:43:54 &	823.0 	 &	3.15e-09 \\
            &	20202510203 &	621137018.0 &	2020Sep07 at 02:03:12 &	865.0 	 &	2.9225e-09 \\
           
00013201082 &	20202621247 &	622126051.0 &	2020Sep18 at 12:47:05 &	452.0 	 &	4.284e-09 \\
           
00013201083 &	20202630129 &	622171746.0 &	2020Sep19 at 01:28:40 &	600.0 	 &	1.52e-09 \\
           
00013201084 &	20202640120 &	622257621.0 &	2020Sep20 at 01:19:55 &	600.0 	 &	4.183e-09 \\
           
00013201085 &	20202650112 &	622343567.0 &	2020Sep21 at 01:12:21 &	300.0 	 &	5.9935e-09 \\
           
00013201086 &	20202660102 &	622429350.0 &	2020Sep22 at 01:02:04 &	1413.0 	 &	2.8653e-09 \\
            &	20202660240 &	622435259.0 &	2020Sep22 at 02:40:33 &	244.0 	 &	1.99e-09 \\
           
00013201087 &	20202670055 &	622515337.0 &	2020Sep23 at 00:55:11 &	900.0 	 &	3.3738e-09 \\
           
00013201088 &	20202680048 &	622601317.0 &	2020Sep24 at 00:48:11 &	900.0 	 &	1.09e-09 \\
            &	20202680104 &	622602517.0 &	2020Sep24 at 01:08:11 &	386.0 	 &	4.797e-09 \\
           
00013201089 &	20202780123 &	623467428.0 &	2020Oct04 at 01:23:22 &	795.0 	 &	3.5332e-09 \\
            &	20202780256 &	623472974.0 &	2020Oct04 at 02:55:48 &	649.0 	 &	3.748e-09 \\
           
00013201090 &	20202790108 &	623552892.0 &	2020Oct05 at 01:07:46 &	291.0 	 &	5.9376e-09 \\
           
00013201092 &	20202802325 &	623719559.0 &	2020Oct06 at 23:25:33 &	1684.0   &	2.4675e-09 \\
           
00013201093 &	20202812317 &	623805468.0 &	2020Oct07 at 23:17:22 &	1755.0 	 &	2.4223e-09 \\
           
00013201094 &	20202830045 &	623897158.0 &	2020Oct09 at 00:45:32 &	300.0 	 &	5.422e-09 \\
            &	20202822329 &	623892555.0 &	2020Oct08 at 23:28:49 &	648.0 	 &	3.5875e-09 \\

\\ \hline \\
\end{tabular}
}
\end{table}

\clearpage

\begin{table}
{\footnotesize
	\caption{List of Swift-BAT survey-mode observations of FRB~180916. (Continued)}\label{Table:obs}
	  \begin{tabular}{lllllllllllllll} \hline\hline

Obs ID  &	 Pointing ID	&	Start time (MET) &	Start time (UTC) &	exposure &	flux upper-lim\\
		&number		&		&year:day:hr:min:sec &(Sec)		&$\funit{}$	 \\ \hline \\

00013201095 &	20202850045 &	624069912.0 &	2020Oct11 at 00:44:46 &	531.0 	 &	3.962e-09 \\
            &	20202842306 &	624063983.0 &	2020Oct10 at 23:05:57 &	1300.0 	 &	2.7107e-09 \\
           
00013201096 &	20202960710 &	625043405.0 &	2020Oct22 at 07:09:39 &	1438.0 	 &	2.7067e-09 \\
            &	20202960540 &	625038245.0 &	2020Oct22 at 05:43:39 &	959.0 	 &	3.256e-09 \\
            &	20202960534 &	625037645.0 &	2020Oct22 at 05:33:39 &	300.0 	 &	5.359e-09 \\
           
00013201097 &	20202970710 &	625129843.0 &	2020Oct23 at 07:10:17 &	1639.0 	 &	2.735e-09 \\
           
00013201098 &	20202980839 &	625221570.0 &	2020Oct24 at 08:39:04 &	1593.0 	 &	2.5897e-09 \\
           
00013201099 &	20202991142 &	625318925.0 &	2020Oct25 at 11:41:39 &	1738.0 	 &	2.517e-09 \\
           
00013201100 &	20203001929 &	625433370.0 &	2020Oct26 at 19:29:04 &	1653.0 	 &	2.126e-09 \\
           
00013201101 &	20203011759 &	625514363.0 &	2020Oct27 at 17:58:57 &	600.0 	 &	3.018e-09 \\
           
00013201102 &	20203131851 &	626554315.0 &	2020Nov08 at 18:51:29 &	300.0 	 &	4.32e-09 \\
            &	20203132200 &	626565653.0 &	2020Nov08 at 22:00:27 &	300.0 	 &	4.849e-09 \\
            &	20203131716 &	626548616.0 &	2020Nov08 at 17:16:30 &	247.0 	 &	4.689e-09 \\
            &	20203132031 &	626560256.0 &	2020Nov08 at 20:30:30 &	187.0 	 &	6.013e-09 \\
           
00013201103 &	20203140731 &	626599904.0 &	2020Nov09 at 07:31:18 &	919.0 	 &	2.765e-09 \\
           
00013201104 &	20203142018 &	626645888.0 &	2020Nov09 at 20:17:42 &	955.0 	 &	2.5645e-09 \\
            &	20203142149 &	626651392.0 &	2020Nov09 at 21:49:26 &	1151.0   &	7.94e-10 \\

00013201111 &	20203440406 &	629179624.0 &	2020Dec09 at 04:06:37 &	1319.0 	 &	2.524e-09 \\
           
00013201112 &	20203450531 &	629271120.0 &	2020Dec10 at 05:31:33 &	1683.0 	 &	2.3245e-09 \\
           
00013201113 &	20203481134 &	629552099.0 &	2020Dec13 at 11:34:32 &	1444.0 	 &	2.415e-09 \\
           
00013201114 &	20203490816 &	629626618.0 &	2020Dec14 at 08:16:31 &	1565.0 	 &	2.519e-09 \\
           
00013201115 &	20203501119 &	629723996.0 &	2020Dec15 at 11:19:29 &	900.0 	 &	3.0374e-09 \\
           
00013201116 &	20203651817 &	631045081.0 &	2020Dec30 at 18:17:34 &	782.0 	 &	3.335e-09 \\
            &	20203651956 &	631050977.0 &	2020Dec30 at 19:55:50 &	646.0 	 &	3.897e-09 \\
           
00013201117 &	20203661959 &	631137568.0 &	2020Dec31 at 19:59:01 &	300.0 	 &	5.841e-09 \\
            &	20203661808 &	631130943.0 &	2020Dec31 at 18:08:36 &	1260.0 	 &	2.591e-09 \\
           
00013201118 &	20210011139 &	631193954.0 &	2021Jan01 at 11:38:47 &	649.0 	 &	3.2846e-09 \\
            &	20210010827 &	631182432.0 &	2021Jan01 at 08:26:45 &	1311.0 	 &	2.5016e-09 \\
           
00013201119 &	20210100240 &	631939239.0 &	2021Jan10 at 02:40:12 &	1764.0 	 &	2.389e-09 \\
           
00013201120 &	20210110233 &	632025217.0 &	2021Jan11 at 02:33:10 &	1766.0 	 &	2.3602e-09 \\
           
00013201121 &	20210122136 &	632180188.0 &	2021Jan12 at 21:36:01 &	1595.0 	 &	2.2733e-09 \\
           
00013201122 &	20210131951 &	632260302.0 &	2021Jan13 at 19:51:15 &	1701.0 	 &	2.40e-09 \\
            &	20210132127 &	632266051.0 &	2021Jan13 at 21:27:04 &	1712.0 	 &	2.20e-09 \\
           
00013201123 &	20210141634 &	632334864.0 &	2021Jan14 at 16:33:57 &	1719.0 	 &	2.4425e-09 \\
           
00013201124 &	20210151938 &	632432312.0 &	2021Jan15 at 19:38:05 &	1711.0 	 &	2.3778e-09 \\
           
00013201125 &	20210160213 &	632455971.0 &	2021Jan16 at 02:12:24 &	300.0 	 &	5.7285e-09 \\
            &	20210160157 &	632455071.0 &	2021Jan16 at 01:57:24 &	300.0 	 &	5.201e-09 \\
           
00013201132 &	20210331559 &	633974372.0 &	2021Feb02 at 15:59:05 &	287.0 	 &	2.7897e-09 \\
           
00013201134 &	20210621452 &	636475946.0 &	2021Mar03 at 14:51:59 &	637.0 	 &	3.995e-09 \\
           
00013201136 &	20210641324 &	636643500.0 &	2021Mar05 at 13:24:33 &	483.0 	 &	1.94e-09\\
           
00013201139 &	20210771309 &	637765766.0 &	2021Mar18 at 13:08:59 &	300.0 	 &	5.444e-09 \\
           
00013201142 &	20210801305 &	638024954.0 &	2021Mar21 at 13:08:47 &	409.0 	 &	4.538e-09 \\
           
00013201143 &	20210811240 &	638109627.0 &	2021Mar22 at 12:40:00 &	1716.0 	 &	2.37e-09 \\
           
00013201145 &	20211091202 &	640526551.0 &	2021Apr19 at 12:02:26 &	212.0 	 &	6.509e-09 \\
           
00013201146 &	20211101131 &	640611085.0 &	2021Apr20 at 11:31:20 &	1658.0 	 &	2.618e-09 \\
           
00013201147 &	20211111124 &	640697088.0 &	2021Apr21 at 11:24:43 &	1635.0 	 &	2.6705e-09 \\
           
00013201148 &	20211121117 &	640783039.0 &	2021Apr22 at 11:17:14 &	1484.0 	 &	2.753e-09 \\
            &	20211121610 &	640800645.0 &	2021Apr22 at 16:10:40 &	678.0	 &	3.474e-09 \\
           
00013201149 &	20211130001 &	640828877.0 &	2021Apr23 at 00:01:12 &	300.0 	 &	4.806e-09 \\
            &	20211131110 &	640869039.0 &	2021Apr23 at 11:10:34 &	1644.0 	 &	2.5765e-09 \\
            &	20211130009 &	640829477.0 &	2021Apr23 at 00:11:12 &	166.0 	 &	6.636e-09 \\
           
00013201150 &	20211141107 &	640955238.0 &	2021Apr24 at 11:07:13 &	1605.0 	 &	2.519e-09 \\

\\ \hline \\
\end{tabular}
}
\end{table}

\clearpage

\begin{table}
{\footnotesize
	\caption{List of Swift-BAT survey-mode observations of FRB~180916. (Continued)}\label{Table:obs}
	  \begin{tabular}{lllllllllllllll} \hline\hline 

Obs ID  &	 Pointing ID	&	Start time (MET) &	Start time (UTC) &	exposure &	flux upper-lim\\
		&number		&		&year:day:hr:min:sec &(Sec)		&$\funit{}$	 \\ \hline \\

00013201151 &	20211151100 &	641041209.0 &	2021Apr25 at 11:00:04 &	1614.0 	 &	2.559e-09 \\
            &	20211150436 &	641018178.0 &	2021Apr25 at 04:36:13 &	1005.0 	 &	2.917e-09 \\
           
00013201153 &	20211290922 &	642244925.0 &	2021May09 at 09:22:00 &	600.0 	 &	3.751e-09 \\
            &	20211291055 &	642250528.0 &	2021May09 at 10:55:23 &	600.0 	 &	3.76e-09 \\
           
00013201154 &	20211300908 &	642330541.0 &	2021May10 at 09:08:56 &	970.0 	 &	2.9134e-09 \\
           
00013201155 &	20211311051 &	642423373.0 &	2021May11 at 10:56:08 &	350.0 	 &	4.7754e-09 \\
           
00013201159 &	20211441041 &	643545682.0 &	2021May24 at 10:41:17 &	278.0 	 &	5.653e-09 \\
            &	20211440911 &	643540283.0 &	2021May24 at 09:11:18 &	460.0 	 &	4.28e-09 \\
           
00013201160 &	20211450555 &	643614931.0 &	2021May25 at 05:55:26 &	582.0 	 &	3.3717e-09 \\
           
00013201161 &	20211580807 &	644746078.0 &	2021Jun07 at 08:07:53 &	1385.0 	 &	2.673e-09 \\
           
00013201162 &	20211590800 &	644832053.0 &	2021Jun08 at 08:00:48 &	1200.0 	 &	2.7623e-09 \\
           
00013201163 &	20211600753 &	644918000.0 &	2021Jun09 at 07:53:15 &	1200.0 	 &	2.908e-09 \\
           
00013201164 &	20211610746 &	645004014.0 &	2021Jun10 at 07:46:49 &	1200.0 	 &	2.9817e-09 \\
           
00013201165 &	20211620740 &	645090007.0 &	2021Jun11 at 07:40:02 &	1200.0 	 &	2.8246e-09 \\
           
00013201166 &	20211630732 &	645175978.0 &	2021Jun12 at 07:32:53 &	1200.0 	 &	2.779e-09 \\
           
00013201167 &	20211640724 &	645261882.0 &	2021Jun13 at 07:24:37 &	1461.0 	 &	2.6162e-09 \\
           
00013201169 &	20211730447 &	646030030.0 &	2021Jun22 at 04:47:05 &	1200.0 	 &	2.7585e-09 \\
            &	20211730510 &	646031530.0 &	2021Jun22 at 05:12:05 &	173.0 	 &	6.483e-09 \\
           
00013201170 &	20211740745 &	646127135.0 &	2021Jun23 at 07:45:30 &	1708.0 	 &	2.1495e-09 \\
           
00013201171 &	20211750620 &	646208442.0 &	2021Jun24 at 06:20:37 &	1041.0 	 &	2.642e-09 \\
           
00013201172 &	20211761548 &	646328943.0 &	2021Jun25 at 15:48:58 &	360.0 	 &	4.869e-09 \\
            &	20211770908 &	646391293.0 &	2021Jun26 at 09:08:08 &	710.0 	 &	3.3605e-09 \\
            &	20211761856 &	646340223.0 &	2021Jun25 at 18:56:58 &	840.0 	 &	3.0734e-09 \\
            &	20211772334 &	646443271.0 &	2021Jun26 at 23:34:26 &	512.0 	 &	4.242e-09 \\
            &	20211760602 &	646293757.0 &	2021Jun25 at 06:02:32 &	1707.0 	 &	2.2164e-09 \\
           
00013201173 &	20211780424 &	646460698.0 &	2021Jun27 at 04:24:53 &	545.0 	 &	3.6787e-09 \\
            &	20211780719 &	646471186.0 &	2021Jun27 at 07:19:41 &	737.0 	 &	3.207e-09 \\
            &	20211790904 &	646563890.0 &	2021Jun28 at 09:04:45 &	853.0 	 &	2.981e-09 \\
           
00013201174 &	20211790535 &	646551329.0 &	2021Jun28 at 05:35:24 &	1114.0	 &	2.835e-09 \\
            &	20211790711 &	646557091.0 &	2021Jun28 at 07:11:26 &	392.0 	 &	4.593e-09 \\
           
00013201175 &	20211890605 &	647417112.0 &	2021Jul08 at 06:05:07 &	1431.0 	 &	4.57e-10\\
           
00013201176 &	20211900554 &	647502881.0 &	2021Jul09 at 05:54:36 &	1702.0 	 &	2.365e-09 \\
           
00013201177 &	20211910729 &	647594963.0 &	2021Jul10 at 07:29:18 &	460.0 	 &	4.218e-09 \\
            &	20211910555 &	647589325.0 &	2021Jul10 at 05:55:20 &	1238.0 	 &	2.5826e-09 \\
           
00013201178 &	20211920542 &	647674932.0 &	2021Jul11 at 05:42:07 &	951.0 	 &	3.031e-09 \\
           
00013201180 &	20211940217 &	647835482.0 &	2021Jul13 at 02:17:57 &	781.0 	 &	3.2184e-09 \\
            &	20211940531 &	647847077.0 &	2021Jul13 at 05:31:12 &	706.0 	 &	3.4086e-09 \\
           
00013201181 &	20211950659 &	647938771.0 &	2021Jul14 at 06:59:26 &	812.0 	 &	3.408e-09 \\
            &	20211950525 &	647933106.0 &	2021Jul14 at 05:25:01 &	597.0 	 &	-8.812 \\
           
00013201182 &	20212060447 &	648881252.0 &	2021Jul25 at 04:47:27 &	391.0 	 &	5.223e-09 \\
            &	20212060621 &	648886885.0 &	2021Jul25 at 06:21:20 &	459.0 	 &	5.304e-09 \\
           
00013201183 &	20212070614 &	648972868.0 &	2021Jul26 at 06:14:23 &	600.0 	 &	4.791e-09 \\
            &	20212070439 &	648967168.0 &	2021Jul26 at 04:39:23 &	875.0 	 &	3.613e-09 \\
           
00013201184 &	20212080431 &	649053124.0 &	2021Jul27 at 04:31:59 &	1139.0	 &	3.385e-09 \\
            &	20212080607 &	649058880.0 &	2021Jul27 at 06:07:55 &	244.0 	 &	7.674e-09 \\
           
00013201185 &	20212090419 &	649138798.0 &	2021Jul28 at 04:19:53 &	1500.0 	 &	2.9306e-09 \\
           
00013201186 &	20212100413 &	649224800.0 &	2021Jul29 at 04:13:15 &	1723.0 	 &	2.727e-09 \\
           
00013201187 &	20212110410 &	649311053.0 &	2021Jul30 at 04:10:48 &	1270.0 	 &	3.1487e-09 \\
           
00013201188 &	20212121021 &	649419943.0 &	2021Jul31 at 10:25:38 &	600.0 	 &	4.162e-09 \\
           
00013201189 &	20212560558 &	653205519.0 &	2021Sep13 at 05:58:34 &	264.0 	 &	6.923e-09 \\
           
00013201191 &	20212600035 &	653531755.0 &	2021Sep17 at 00:35:50 &	300.0 	 &	6.383e-09 \\
            &	20212600050 &	653532655.0 &	2021Sep17 at 00:50:50 &	300.0 	 &	5.919e-09 \\
           
00013201199 &	20212760010 &	654912618.0 &	2021Oct03 at 00:10:13 &	1665.0 	 &	2.19e-09 \\
           
00013201200 &	20212770003 &	654998591.0 &	2021Oct04 at 00:03:06 &	300.0 	 &	4.91e-09 \\

\\ \hline \\
\end{tabular}
}
\end{table}

\clearpage

\begin{table}
{\footnotesize
	\caption{List of Swift-BAT survey-mode observations of FRB~180916. (Continued)}\label{Table:obs}
	  \begin{tabular}{lllllllllllllll} \hline\hline 

Obs ID  &	 Pointing ID	&	Start time (MET) &	Start time (UTC) &	exposure &	flux upper-lim\\
		&number		&		&year:day:hr:min:sec &(Sec)		&$\funit{}$	 \\ \hline \\

00013201201 &	20212771745 &	655062358.0 &	2021Oct04 at 17:45:53 &	1325.0 	 &	2.526e-09 \\
           
00013201202 &	20212780131 &	655090278.0 &	2021Oct05 at 01:31:13 &	1665.0 	 &	2.1616e-09 \\
           
00013201203 &	20212912230 &	656289186.0 &	2021Oct18 at 22:33:01 &	477.0 	 &	4.265e-09 \\
            &	20212912218 &	656288286.0 &	2021Oct18 at 22:18:01 &	600.0 	 &	1.54e-09\\
           
00013201204 &	20213072113 &	657666951.0 &	2021Nov03 at 21:15:46 &	192.0 	 &	6.118e-09 \\
           
00013201205 &	20213082235 &	657758161.0 &	2021Nov04 at 22:35:56 &	300.0 	 &	5.638e-09 \\
           
00013201207 &	20213102215 &	657929752.0 &	2021Nov06 at 22:15:47 &	1200.0 	 &	2.9644e-09 \\
           
00013201210 &	20213232044 &	659047485.0 &	2021Nov19 at 20:44:40 &	900.0 	 &	3.343e-09 \\
           
00013201211 &	20213242041 &	659133687.0 &	2021Nov20 at 20:41:22 &	1200.0 	 &	2.875e-09 \\
           
00013201214 &	20213362039 &	660170354.0 &	2021Dec02 at 20:39:09 &	1369.0 	 &	2.3474e-09 \\

00013201215 &	20213372031 &	660256279.0 &	2021Dec03 at 20:31:14 &	1304.0 	 &	2.289e-09 \\
           
00013201216 &	20213382036 &	660342979.0 &	2021Dec04 at 20:36:14 &	1364.0 	 &	2.1998e-09 \\
           
00013201217 &	20213392016 &	660428224.0 &	2021Dec05 at 20:16:59 &	1679.0 	 &	2.0684e-09 \\
           
00013201218 &	20213402009 &	660514188.0 &	2021Dec06 at 20:09:43 &	1694.0 	 &	2.1318e-09 \\
           
00013201219 &	20213412003 &	660600200.0 &	2021Dec07 at 20:03:15 &	1663.0 	 &	2.141e-09 \\
           
00013201220 &	20213421955 &	660686158.0 &	2021Dec08 at 19:55:53 &	1685.0 	 &	2.207e-09 \\
           
00013201221 &	20213591835 &	662150105.0 &	2021Dec25 at 18:35:00 &	1018.0 	 &	2.785e-09 \\
           
00013201224 &	20220041732 &	663010338.0 &	2022Jan04 at 17:32:13 &	1005.0 	 &	3.1497e-09 \\
            &	20220041855 &	663015305.0 &	2022Jan04 at 18:55:00 &	898.0 	 &	3.3017e-09 \\
           
00013201225 &	20220051722 &	663096183.0 &	2022Jan05 at 17:22:58 &	960.0 	 &	3.1152e-09 \\
            &	20220051847 &	663101271.0 &	2022Jan05 at 18:47:46 &	852.0 	 &	5.12e-09\\
           
00013201226 &	20220061700 &	663181256.0 &	2022Jan06 at 17:00:51 &	1147.0 	 &	3.1096e-09 \\
            &	20220061840 &	663187256.0 &	2022Jan06 at 18:40:51 &	547.0 	 &	4.237e-09 \\
           
00013201227 &	20220071831 &	663273119.0 &	2022Jan07 at 18:31:54 &	1624.0 	 &	7.94e-10 \\
           
00013201228 &	20220081823 &	663359039.0 &	2022Jan08 at 18:23:54 &	1684.0 	 &	2.4566e-09 \\
           
00013201229 &	20220091815 &	663444962.0 &	2022Jan09 at 18:15:57 &	1681.0 	 &	2.415e-09 \\
           
00013201230 &	20220101807 &	663530826.0 &	2022Jan10 at 18:07:01 &	1677.0 	 &	2.385e-09 \\

\\ \hline \\
\end{tabular}
}
\end{table}

\clearpage
                
                
\acknowledgements
The material is based upon work supported by NASA under award number 80GSFC21M0002. 
ET acknowledges support from NASA grant 80NSSC18K0429. MN is supported by the European Research Council (ERC) under the European Union’s Horizon 2020 research and innovation programme (grant agreement No. 948381) and by a Fellowship from the Alan Turing Institute.

\bibliographystyle{aasjournal}
\bibliography{mybib}

\end{document}